# Exciton-Polaritons Dynamics of a Monolayer Tungsten Disulphide ($WS_2$) Coupled to a Semiconductor Microcavity

Mohamed Herira[1], Hela Boustanji[2] and Sihem Jaziri[1,2]

[1]*Laboratoire de Physique des Matériaux, Faculté des Sciences de Bizerte, Université de Carthage, 7021 Jarzouna, Tunisia*

[2]*Laboratoire de Physique de la Matière Condensée, Département de Physique, Faculté des Sciences de Tunis, Université Tunis el Manar, Campus Universitaire, 1060 Tunis, Tunisia*

## Abstract

We present a theoretical model that allows us to describe the dynamics of exciton polaritons in the strong-coupling regime in a monolayer $WS_2$ based semiconductor microcavity. Numerical simulations using Boltzmann equations give an overall description of polariton kinetics over a temperature range of 130–230K, under nonresonant excitation. Here, only the scattering rate via optical phonons (LO) from the upper (UP) towards lower (LP) polariton branches is considered. According this model we show the importance of the radiative lifetime of the polaritonic states relative to polariton relaxation rate. Our results show as the cavity detuning changes from negative (130K) to positive (230K) values, the UP branch can be tuned from a more excitonlike to a more photonlike state at small wave vector **k**. We deduce that the UP states have a faster lifetime and the polariton relaxation time into the LP energy states becomes very efficient. Thus, the UP occupation starts to decrease, and we observe a simultaneous increase in the occupation of the LP branch. Furthermore, the polariton states are less occupied for small excitation pump rate, the calculated behavior is linear. In the high pumping regime, we observe the nonlinear behavior of cavity polaritons and occupation factors much larger than unity are reached.

.

**Keys words:** Monolayer Tungsten Disulphide ($WS_2$), Semiconductor Microcavity, strong, coupling regime, Polaritons dynamics, Boltzmann equations, scattering rates, Optical phonons (LO), radiative recombination rate.

I. Introduction:

Coherent properties of exciton-polaritons in many semiconducting materials have recently attracted great interest [1-5]. An exciton-polariton is a quasiparticle resulting from the strong coupling of electron-hole pairs (excitons in a semiconductor) and cavity photon [3]. The optical properties of excitons can be significantly changed when coupled to photonic cavities modes [2, 3, 6]. Different cavity structures such as the distributed Bragg Reflector (DBR) have been utilized to manipulate the optical properties of polaritons. Cavity polaritons have been demonstrated in a wide range of materials [7-9]. The microcavity polaritons were demonstrated at cryogenic temperatures, in both CdTe and GaAs microcavities due to the small binding energy of excitons [7-9]. Many material systems have now shown polaritons at room temperature including wide-bandgap semiconductors such as GaN, ZnO …. [10-12], organic material [13,14] and more recently transition metal dichalcogenides (TMDs) [1,3]. The strongly confined excitons and the large binding energy of TMDs monolayers make them ideal candidates to allow the strong light-matter interactions under ambient conditions. Exciton-polaritons in TMDs materials have been observed experimentally [11–13]. By embedding monolayer of molybdenum disulphide ($MoS_2$) into distributed Bragg reflector (DBR) mirrors, X. Liu et al. first reported room temperature 2D exciton polaritons with Rabi splitting of 46,3 meV is observed [15]. Part-light–part-matter polariton eigenstates are observed in molybdenum diselenide/hexagonal boron nitride (MoSe$_2$/hBN) hetero-structure tunable micro-cavity with a splitting of 20 meV [16]. Exfoliated Tungsten diselenide($WSe_2$) monolayers have a strongly enhanced luminescence under ambient conditions [17], which suggests their suitability for room-temperature. Also, the monolayer tungsten disulphide ($WS_2$) have shown a strong interaction with light and robust excitons at room temperature [3]. In microcavity based on monolayer $WS_2$, the reflectivity spectra have shown two pronounced modes corresponding to the Upper (UP) and Lower (LP) polariton branches appear at small incidence angles. Moreover the LP exhibits the stronger emission with a maximum peak around the lowest polariton state [3]. To better understand the emission properties of the monolayer $WS_2$ embedded microcavity, we have to analyze the reflectivity spectra and studying the dynamic process. In parallel with the experimental progress, theoretical studies and numerical simulations are developed for describing the dynamical nature of a polariton condensate in TMDs.

Boltzmann and Gross-Pitaevskii equations have been well adapted to describe the dynamics of microcavity polaritons. On one hand, approaches based on the Gross-Pitaevskii equation are successful in explaining of recent experiments [18-20], but they assume global coherence and therefore cannot describe scattering process assisted by phonon. On the other hand, the semiclassical Boltzmann equation approaches are widely investigated under resonant or nonresonant excitation [21-25]. Tassone and Yamamoto, [26] have studied the polariton relaxation, which is described by solving a set of semiclassical Boltzmann equations. Porras et al [22] followed Tassone and Yamamoto's 1999 model with some simplifications to the numeric. Tassone et al. [27] have shown that the polaritons dynamics depends on several parameters, the excitation of the system, relaxation toward the lowest state mediated by exciton-phonon interaction and finally the radiative lifetime of polaritons. Exciton-polariton scattering is identified as the most important mechanism for relaxation towards the lower polariton branch.

The dynamics of exciton-polaritons can be primarily understood by the interplay between radiative recombination rate and scattering processes. The former radiative lifetime is determined by the lifetime of photon and exciton modes with appropriate fractional values for each mode. Therefore, this time is very short for photon-like polariton and much longer for exciton-like polariton. In addition, a different scattering mechanism is proposed based on the interactions of exciton-phonon, polariton-polariton and exciton-exciton. However, different works have demonstrated that polariton-polariton scattering is only efficient for resonant optical pumping of the cavity [28]. Two dominant exciton-polariton relaxation processes have been extensively studied, one is due to exciton-acoustic phonon (LA) [27-29] and the other due to exciton-optical (LO) interaction [30-32]. Furthermore, LA phonons cannot assist in the polariton scattering process for small in-plane wave vectors [28]. Tassone et al. observed that large-momentum polaritons are efficiently relaxed to lower energy regions through the emission or absorption of LA phonons. However, they also found that phonon scattering is significantly reduced towards low wave vectors values of the LP branch [27,33].The inclusion of the scattering with LO-phonons may speed up the relaxation kinetics. L. Orosz et al. [34] have studied polariton relaxation mechanisms mediate by an LO phonons in a ZnO-based microcavity. O. Jamadi et al . [35] have observed and discussed the polariton scattering through LO phonons as a function of temperature and detuning in a GaN and ZnO microcavities.

Motivated by the experimental results of the reflectivity spectra shown by Xiaoze Liu et al. [3], which demonstrate the evidence of strong light-matter coupling and formation of microcavity polaritons in a monolayer $WS_2$ based microcavity. The aim of this paper is to analyze in detail the emission of the two reflectivity modes observed at **k**=0 over a temperature range of (110–230 K). Therefore, we present a theoretical study of the evolution of the polariton population as a function of temperatures. We focus on the impact of the shape of the UP and LP branches dispersion curves on the relaxation efficiency. To this purpose, let us first recall the polariton dispersion in Sec.II. Then, we simulate the polariton relaxation in a $WS_2$ microcavity using semiclassical Boltzmann equations under non-resonant pumping rate. To introduce the rate equations approach developed by Tassone et al [26, 27], we first investigate in Sec.III the polariton scattering with optical phonons. We then discuss in Sec.IV the numerical integration of the rate equations and the temporel evolution of the polariton distribution.

## II. Analysis of the strong coupling regime for various temperatures in a monolayer $WS_2$ based semiconductor microcavity:

II.1 Light–matter interactions in the strong coupling regime: Polariton dispersion energies.

The microcavity structure studied theoretically in this paper and the experimental method are described in Ref. [3]. In the experiment, polaritons are created in a monolayer $WS_2$ based semiconductor microcavity. The microcavity structure consists of a bottom distributed Bragg reflector (DBR) mirror grown on a substrate, a cavity which contains the active medium monolayer of $WS_2$ sandwiched between $HSQ$ and $HSQ/Al_2O_3$ layers and a top DBR mirror [3]. The DBRs consists multiple periods of alternative $SiO_2$ and $Si_3N_4$ layers. A schematic of a planar microcavity is depicted in FIG. 1(a).

In the following, we discuss the coupling between the cavity mode and excitonic states in a monolayer $WS_2$ using the coupled oscillator model. Due to the in-plane momentum conservation, the interaction is nonzero only for photon and exciton states have a wave vector within the light cone. Because of the angle dependence of the reflectance of the Bragg mirrors, the photon confinement is efficient only near the center of the Brillouin zone. The confinement of light gives the cavity mode a parabolic dispersion in the plane perpendicular to the confinement direction, which can be written as $E_{cav}(\boldsymbol{k}) = \frac{\hbar c}{n_{eff}} \sqrt{\frac{\Pi^2}{L_c^2} + \boldsymbol{k}^2}$, where $L_c$ is cavity length. Only inside this region $|\mathbf{k}| < \mathbf{k_0} = \frac{n_{eff} \mathbf{E_{ex}(k=0)}}{\hbar c}$ the coupling between photon-exciton is strong, where $n_{eff} = \sqrt{\varepsilon_{eff}}$ is the effective refractive index of the cavity layer.

Here $\varepsilon_{eff}$ is the effective dielectric constant, because the monolayer $WS_2$ material is surrounded by an environment with dielectric constants $\varepsilon_1$ (top) and $\varepsilon_2$ (down). $E_{ex}(\mathbf{k}=0)$ denote the energy of exciton at the zero in-plane momentum $\mathbf{k}$. The in-plane dispersion relation for excitons reads $E_{ex}(\mathbf{k}) = \frac{\hbar^2 \mathbf{k}^2}{2M_X} - E_b + E_g(T)$, here $M_X$ is the free in-plane excitonmass, which is the sum of electron and hole effective masses. The following parameters have been considered for the calculations, effective electron and heavy hole masses $m_e = m_h = 0.34\, m_0$, with $m_0$ the free electron mass. To calculate the exciton binding energy $E_b$, we use the hydrogenic model employs an effective mass Hamiltonian. Here, the interaction potential $V_{2D}(\mathbf{r})$ is given by the 2D screened electrostatic interaction potential derived by Keldysh [36, 37]. We find an exciton binding energy of $E_b = 205\, meV$, which is discussed in Appendix A. Alexey Chernikov et al. [37] have determined experimentally the energies of the ground and the excited excitonic states optical transition in monolayer $WS_2$. From the spectra, they have observed a large exciton binding energy of 0.32 eV. The temperature dependence of energy gap is given by $E_g(T)$ (see Appendix B for details).

The Wannier-Mott excitons in $WS_2$ exhibits a large binding energy that enable the tuning of the exciton part over a large temperature range [3, 36]. Due to the emergence of dark excitons and trions at low temperature, our study is limited at temperatures higher than 110K, where the exciton peak dominates [3,36] (see Appendix B). In this context, we study the evolution of the strongly coupled system by varying its temperature.

The polariton dispersion can be modeled using the coupled oscillator model

$$\begin{pmatrix} E_{ex} + i\hbar\Gamma_{ex} & V_A \\ V_A & E_{cav} + i\hbar\Gamma_{cav} \end{pmatrix} \begin{pmatrix} X_\mathbf{k} \\ C_\mathbf{k} \end{pmatrix} = E \begin{pmatrix} X_\mathbf{k} \\ C_\mathbf{k} \end{pmatrix}$$

From the reservoir, excitons relax by acoustic phonon towards the light-cone with non-radiative rate $\Gamma_{ex}$. We can calculate this rate using the Fermi Golden rule and the optical deformation potential (More details are provided in Appendix. C). $\Gamma_{cav}$ is the half width half maximum of cavity photon [3]. The coefficients $X_\mathbf{k}$ and $C_\mathbf{k}$, are called the exciton and cavity Hopfield coefficients. Thus, can be given in terms of the detuning $\Delta = E_{cav}(k) - E_{ex}(k)$ and the interaction potential $V_A$ (see Appendix. C). E are the eigen energies of the Hamiltonian at each wave vector $\mathbf{k}$. Here, the two polariton modes are described by the real parts of the energy eigen-value and can be expressed as a function of detuning energy. The eigen-states of the coupled system are a linear superpositionof exciton$|\Psi_{ex,\mathbf{k}}\rangle$and photon $|\Psi_{c,\mathbf{k}}\rangle$modulated by the Hopfield coefficients$X_\mathbf{k}$ and $C_\mathbf{k}$.

Dispersion curves of the LP and UP polariton branches are shown in FIG. 1(b) and (c) at temperatures 130 and 230K.Comparison of FIGs.1 (b) and (c) shows, an appreciable redshifts in exciton energy as a function of temperature from 2.069 (130K) to 2.042eV (230K). Thus, the cavity detuning changes from negative -16meV to positive 11meV as the temperature increase. The polariton branches were modified significantly due to the various cavity detunings. The energies and shapes of the polariton dispersion curves depend strongly on $\Delta$. As seen in FIGs. 1(b) and (c), the lower polariton is more photon-like and the upper polariton is more exciton-like for $\Delta < 0$, and the lower polariton is more exciton-like and the upper polariton is more photon-like when $\Delta > 0$.

The second significant property is that the composition (percentage of the excitonic and photonic fractions) of polaritons changes dramatically along the dispersion.The Hopfield coefficients for the LP branch are plotted in FIG. 1(d) and (e) for various temperatures. The Hopfield coefficients show that the LPs states can be tuned from a more photon-like, to a more exciton like state at small wave vectors. Xiaoze Liu et al. [3] have shown that the large contrast between the UP and LP states in the PL distribution, indicates LP is more cavity-like and, thus, has a much faster decay rate. At resonance, both branches of polariton states are exactly half photon half exciton. From FIG. 1(b), we shows at $\mathbf{k}$= 0 the energy splitting between the two polaritons branches is approximately equals to $\hbar\Omega_{rabi} = 39.98$meV (As shown in Appendix C), which yields a coupling strength $V_A = 19\ meV$. From the reflectivity spectra, the strength of radiative coupling between the exciton and cavity field is around $\hbar\Omega_{rabi} = 39$meV at 130K [3].

In the microcavity structures, the coherent coupling depends on the oscillator strength of the exciton transition and the quality factor of the cavity []. The coherent coupling is represented by a large Rabi splitting and reduced polariton linewidth []. More importantly, to maintain strongly coupled exciton-polaritons with large Rabi splitting, this requires a small cavity length and large quality factor. The strong coupling regime can only be achieved if the coupling between light and matter is larger than all the losses. The quality of microcavities depends on the lifetime of the cavity photons. Moreover, the quality factor and the lifetime of polaritons are also important for the polariton relaxation mechanisms.

II.2 Temperature effects on the polaritons radiative lifetime:

The polaritons dynamics are determined by the competition between the scattering processes and the radiative recombination rate. The relative efficiencies of these processes depend on the photonic or excitonic character of the UP and LP branches states. In the following, we calculate the radiative recombination rates $\frac{1}{\tau_{UP(LP)}^{rad}(\mathbf{k})}$ for UP and LP modes, respectively [3].The lifetime of the polaritons is also determined by the Hopfield coefficient as:

$$\frac{1}{\tau_{UP}^{rad}(\mathbf{k})} = \frac{|C_\mathbf{k}|^2}{\tau_{ex}^{rad}} + \frac{|X_\mathbf{k}|^2}{\tau_{cav}^{rad}} \quad and \quad \frac{1}{\tau_{LP}^{rad}(\mathbf{k})} = \frac{|X_\mathbf{k}|^2}{\tau_{ex}^{rad}} + \frac{|C_\mathbf{k}|^2}{\tau_{cav}^{rad}} \quad (1)$$

In our study, the cavity photon lifetime $\tau_{cav}^{rad}$ is taken to be 1ps [3], whereas the radiative lifetime of the neutral exciton in a monolayer $WS_2$ is estimated to be 226ps, which is comparable to those in other 2D TMD semiconductors at low temperatures [38].In monolayer $MoS_2$, the radiative lifetime of exciton is in the range of 0.18–0.30 ps [39]. These lifetimes are shorter than the one measured in III-V semiconductor quantum wells. For wide $GaAs/Ga_{1-x}Al_xAs$ quantum wells, L. C. Andreani et al. [40] have shown the lifetime of the lowest exciton is about 25 ps at zero in-plane wavevector. These results reveal that the monolayer $WS_2$ is characterized by a long radiative lifetimes of excitons.

The lifetimes of thermally distributed excitons in TMDs depend linearly on the temperature and can be in the nanosecond range near room temperature. Theoretical calculations indicate that the PL of TMDs lasts only several picoseconds at cryogenic temperatures (5K) and several nanoseconds at room temperature [41].To determine the average radiative lifetime $\tau_{ex}$ inmonolayer $WS_2$, we use the following relation $\tau_{ex} = \tau_{ex}(0)\frac{3}{4}(\frac{E_{ex}(\mathbf{k}=0)}{4M_X c^2})^{-1}k_B T$ [38,39], here $\tau_{ex}(0)$ is the computed radiative lifetime at 0 K, $c$ is the speed of light, $k_B$ is the Boltzmann constant. At relatively high temperature 230K, we find long-lived excitons 3.9ns which isof great importance to allow exciton polaritons condensation. The resulting radiative lifetimes of microcavity polaritons as a function of the exciton momentum are shown inFIG.1 (f) and (g).

In this system, the radiative lifetimes of the LP and UP modes are of the order of a few ps. The raditaive rate vanishes outside the radiative region $|\mathbf{k}| < \mathbf{k_0}$. In quantum wells only the radiative region of the exciton dispersion is involved in the luminescence processes [42]. In the temperature 130 K, we clearly observe in FIG.1 (f) the LP states have a faster lifetime of the order of1.6psat the zero in-plane momentum $\mathbf{k}$. The larger photonic character of the polaritons with decreasing in-plane momentum significantly reduces the polariton lifetime. From the previous results, we note the exciton radiative lifetime is much longer than the cavity photon lifetime. Therefore, we show that the photon-like polariton lifetime is generally smaller than the more exciton-like polariton lifetime. While at positive detuning, the lifetime of the LP polariton is then dominated by the exciton lifetime (see FIG.1 (g)). Comparison of FIGs.1(f) and (g) shows that the UP branch becomes more photon-like for small value of $\frac{k}{k_0}$, which tends to decrease the UP leakage lifetime.

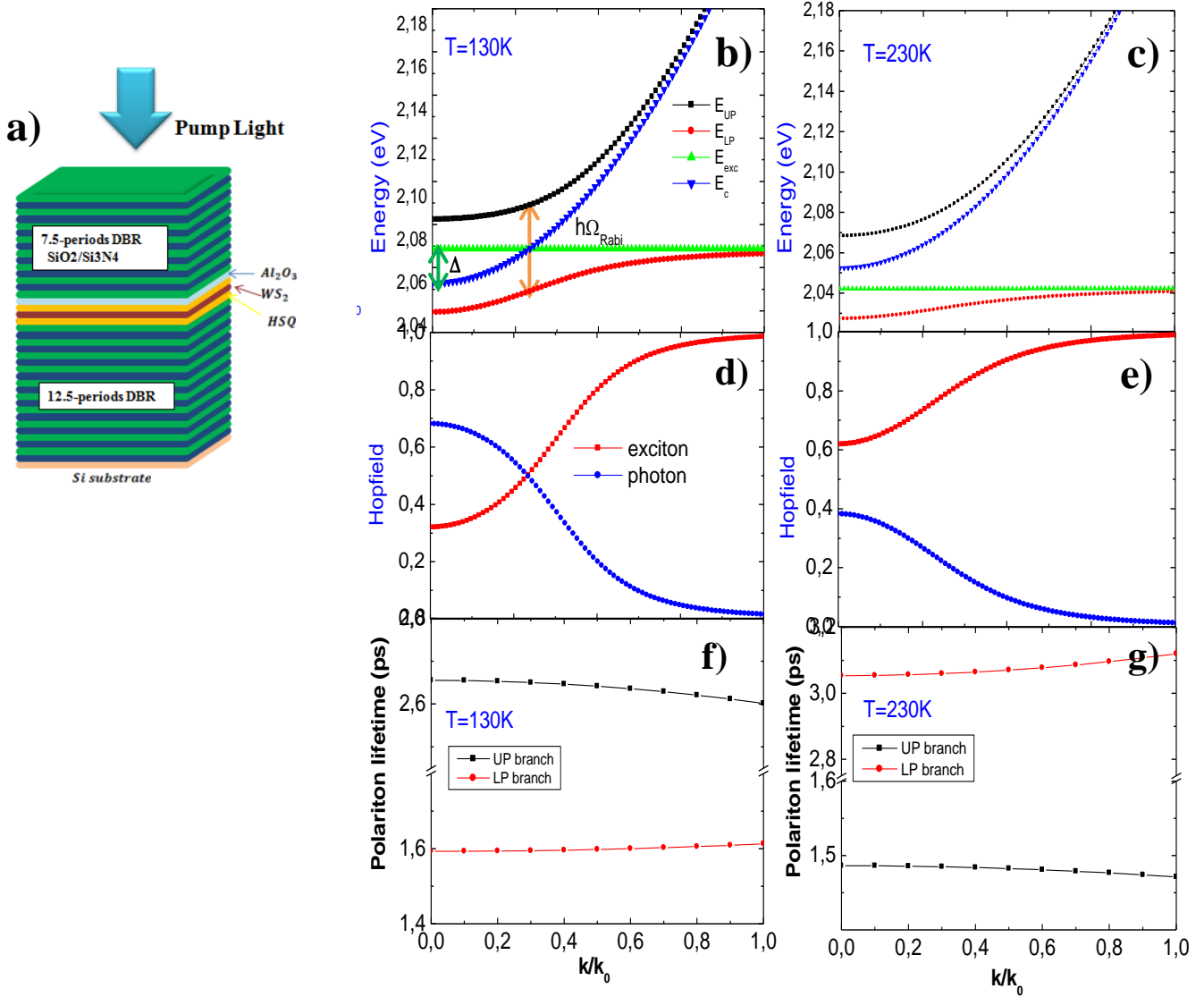

FIG.1: (a) Schematics structure of the microcavity polariton consisting of an active $WS_2$ thin layer sandwiched between distributed Bragg reflectors (DBRs) on the top and bottom [3]. Dispersion curves of polariton modes at temperatures (b) 130 and (c) 230K, which corresponds to a cavity detunings $\Delta = -16$ and 11meV, respectively. (d) and (e) show the corresponding Hopfield coefficients representing the photon and exciton fractions in the LP modes. (f) and (g) Microcavity polariton radiative lifetime as a function of $\frac{k}{k_0}$.

In the strong-coupling regime, the Rabi splitting decreases with increasing temperature. In angle-resolved PL measurements, Xiaoze Liu et al. [3] have observed dispersion of UP and LP polariton branches with Rabi splitting varies from 39 to 36meV at temperature 130 and 230K, which is the energy of an LO-phonon. This means that the lower branch is populated thanks to the scattering between upper polaritons and LO-phonons. Moreover the results of the PL distribution [3], have demonstrated that the polariton state are stable after the scattering and thermal relaxation process. To understand how the polariton states are populated by thermal relaxation processes from the nonresonantly pumped exciton reservoir, we have to study the dynamics of polariton-LO phonon scattering. The following sections describe how these scattering rates are calculated.

### III. Polariton scattering with optical phonons:

In this section, we are interested in calculating the polariton scattering rate induced by the emission of an optical phonon. We demonstrate the role of longitudinal optical (LO)-phonon to accelerate polariton relaxation. The concept of utilising LO-phonon emission was introduced by Imamoglu et al. [31]. The scattering coefficients for both processes $W^{LO}_{UP,k \to LP,k'}$ and $W^{LO}_{UP,k' \to LP,k}$ are computed using the Fermi's Golden Rule. The scattering process discussed here will be used in the following (section. IV). The polariton-phonon model is based on the exciton-phonon interaction, which itself is based on the Frôhlich interaction Hamiltonians. The matrix element has been calculated in Appendix D. The influence of the detuning and the temperature on the cavity polaritons relaxation rate is analyzed.

In FIG.2, we present the results of the total scattering times $\tau_{LO}$ and $\tau'_{LO}$ for different temperatures. At first sight, these scattering times depend on the microcavity structure parameter such as the detuning. Note again from FIG.1(f) the lifetime of LP branch is inversely proportional to the photon fraction in the central region of the dispersion, so it increases once the detuning becomes positive due to the larger excitonic fraction of LP branch (see FIG.1(e)). P. Senellart et al. [43] have shown when going toward a photon-like polariton, interaction with phonons becomes less efficient. As one sees in FIG.1(f) and FIG.1(g), the polariton lifetime is always longer than the relaxation time as shown in FIG.2(a) and FIG.2(b);this leads to an efficient relaxation of the polariton. FIG.2(a) and FIG.2(b) display a comparison between the relaxation times $\tau_{LO}$ and $\tau'_{LO}$, for optical-phonon scattering. One may notice from FIG.2(a) that the scattering time between the states **k** and **k'** is 1.6ps at the lowest **k** state, then decreases to 1.37ps. Another remarkable result, this scattering time is longer than the mean scattering time between the states **k'** and **k** ranges between 1.04 and 1.37ps. From this, we can predict that the increase of the scattering rate means that the relaxation toward the lower branch (the lowest **k** state) becomes very efficient.

Let us now discuss polariton scattering as a function of the exciton photon detuning. We remark from the scattering time shown in FIG.2(a) and FIG.2(b), that increasing the temperature reduces the stimulated scattering time when going from negative to positive value of detuning. The combinations of fast scattering and of a short lifetime of UP branch are responsible for enhancing the polariton distribution.

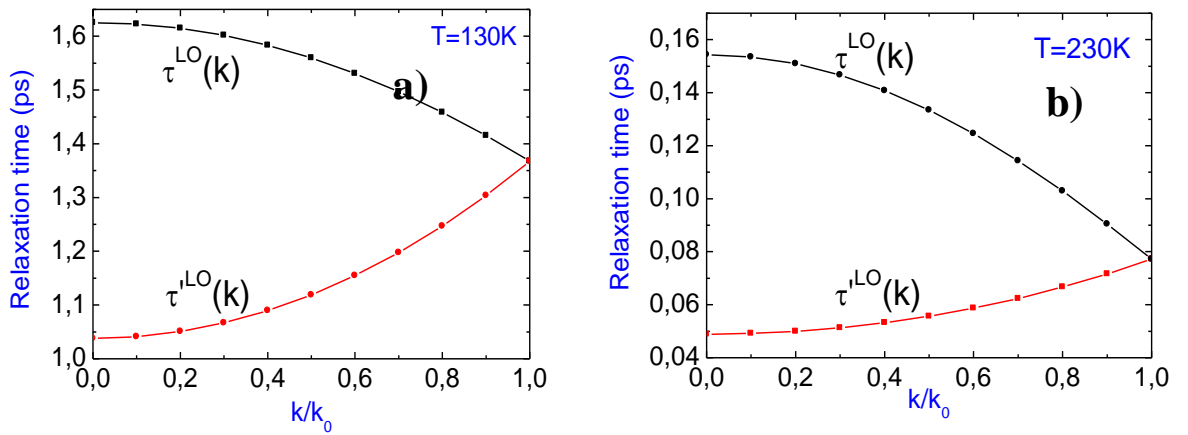

FIG.2. The total scattering time from an initial state $|UP_\mathbf{k}\rangle$ to all final states $|LP_{\mathbf{k'}}\rangle$ (or $|UP_{\mathbf{k'}}\rangle$ to all final states $|LP_\mathbf{k}\rangle$) due to longitudinal optical-phonon emission at temperature (a) 130 and (b) 230K

In the following we will elucidate the emission of the two reflectivity modes appears at **k**=0 [3]. Therefore, we will treat the main equations which describe the polariton kinetic due to exciton-phonon interaction. Our final aim is to derive the semiclassical Boltzmann equations for the population distribution.

## IV. Kinetic models for microcavity polaritons :

In order to understand the time evolution of the polarition distribution, we consider that the microcavity structure is nonresonantly pumped by a pulsed laser excites at high energy 2.43eV. We choose that the duration of the excitation pulse is around 200ps, which is much longer than the polariton lifetime [44]. We write now a set of rate equations for $n_{UP(LP)}$

(occupation numbers in the upper (lower) polariton branch respectively. The rate equations are introduced by using the scattering rates and the total radiative recombination rates. In this case the quantum Boltzmann time evolution equations are found to be:

$$\begin{cases} \dfrac{dn_{UP,\mathbf{k}}}{dt} = P_{\mathbf{k}} - n_{UP,\mathbf{k}} \sum_{\mathbf{k}'} W^{LO}_{UP,\mathbf{k} \to LP,\mathbf{k}'} \left( n_{LP,\mathbf{k}'} + 1 \right) - \dfrac{n_{UP,\mathbf{k}}}{\tau^{rad}_{UP}(\mathbf{k})} & (2) \\ \dfrac{dn_{LP,\mathbf{k}}}{dt} = \left( n_{LP,\mathbf{k}} + 1 \right) \sum_{\mathbf{k}'} W^{LO}_{UP,\mathbf{k}' \to LP,\mathbf{k}} n_{UP,\mathbf{k}'} - \dfrac{n_{LP,\mathbf{k}}}{\tau^{rad}_{LP}(\mathbf{k})} & (3) \end{cases}$$

The generation term describing the optical pumping rate is given by $P_{\mathbf{k}} = \dfrac{qI_{pump}S}{E_{laser}}$, here the quantum efficiency q is about 10% [45], $I_{pump}$ is the optical pump intensity and, S is the sample surface, $E_{laser}$ is the laser energy. The term $(n_{LP,\mathbf{k}} + 1)$ result from the bosonic character of polaritons, and represent stimulated emission process. $W^{LO}_{UP,\mathbf{k} \to LP,\mathbf{k}'}$ ( $W^{LO}_{UP,\mathbf{k}' \to LP,\mathbf{k}}$) describes the scattering rate between the states **k** and **k'** (or between the states **k'** and **k**), respectively. We limit to the contribution of optical-phonons only, which are introduced through the Frôhlich interaction Hamiltonians. $\tau^{rad}_{UP(LP)}(\mathbf{k})$ is the radiative lifetime of polaritons, which are calculated in Sec.II.2. We solve the rate Eq. (2) and (3) by direct numerical integration using *runge-kutta 4* methods. In the numerical calculations we need to discretize the **k**-space. In a finite volume calculation, the **k**-space is uniform. We solve the above rate equations (2) and (3) by using a grid that is uniformly spaced in energy with large number of points in the region from 0 to $10\mathbf{k_0}$.

If we wish to investigate how the polariton states are populated by scattering processes. We need to evaluate the expected evolution of the pump rate $P_{\mathbf{k}}$ as a function of temperature and is summarized for two detuning's.

IV.1. Temperature effects on the optical pumping rate:

From the stationary solution of these nonlinear Equations (2) and (3), we can calculate the threshold pumping intensity $I^{th}_{pump}$ and the threshold optical pumping rate $P^{th}_{\mathbf{k}}$ for a given temperature T.

The steady-state solution $n_{UP,\mathbf{k}}$ (Eq.2) gives the threshold pumping intensity $I^{th}_{pump} = \dfrac{E_{laser}}{q} \dfrac{n^{th}_{UP,\mathbf{k}}}{S} \left( \dfrac{1}{\tau_{LO}} + \dfrac{1}{\tau^{rad}_{UP}} \right)$ versus the threshold pump density $\dfrac{n^{th}_{UP,\mathbf{k}}}{S}$. Thus, the polariton distribution $n_{UP,k}^{th}$ in the steady-state regime will be given by the (Eq. 3)  $\frac{dn_{LP,k}}{dt} = 0$, by taking into account  $n_{LP,k} \ll n_{UP,k}$  and  $n_{LP,k}$  is neglected with respect to 1.  This gives an threshold pump intensity of  $I_{pump}^{th} = 6.03 \ 10^3 \ W/cm^2$ , which corresponds to a threshold pump rate  $P_k^{th} = 1.76 10^{11} s^{-1}$ at temperature  230K.

From the PL measurement [3], the UP PL is much weaker than the LP branch at 130K, so the PL intensity at the higher photon energy panel is magnified by 10 times. While at 230K, The UP intensity is comparable with the LP intensity, so the PL intensity is magnified by 3 times. From this, we can deduce that  the threshold pump rate could be lower at 130K with respect to 230K. We show a linear decrease in $P_k^{th}(130K) = 0.58 \ 10^{11} s^{-1}$, this feature clearly indicates the threshold pump rate dependence on temperature. We can note also the threshold pump rate increases when detuning going toward positive value which can be ascribed to relaxation effect. Jacques Levrat et al. [44] have investigated the evolution of the polariton condensation threshold $P_k^{th}$ over a wide range of temperatures (4–340 K) and exciton-cavity photon detunings (−120–0 meV) in a multiple quantum-well GaN-based microcavity. In both theory and experiment, the results have shown that  $P_k^{th}$ first decreases and then starts increasing again when going toward positive detuning values [44]. The results have shown for a given temperature, the threshold first decreases versus detuning, passes through a minimum and then increases for detunings closer to zero, which can certainly ascribed to  fast relaxation kinetics effects [44].

In the following section, we show the results of the integration of the rate equations over the in plane wave vector, for two different temperature values and excitation intensities.

IV. 2 Temporel evolution of the population $n_{UP}$ and $n_{LP}$ :

Once the scattering rates and the radiative lifetime are known**,** we have integrated numerically the rates equations over the in plane wave vector to obtain the time evolution of the population $n_{UP}$ and $n_{LP}$ . FIG.3 shows the calculated occupation of the UP and LP states for two different pump rate, below and above threshold as a function of temperature T=130 and 230K.

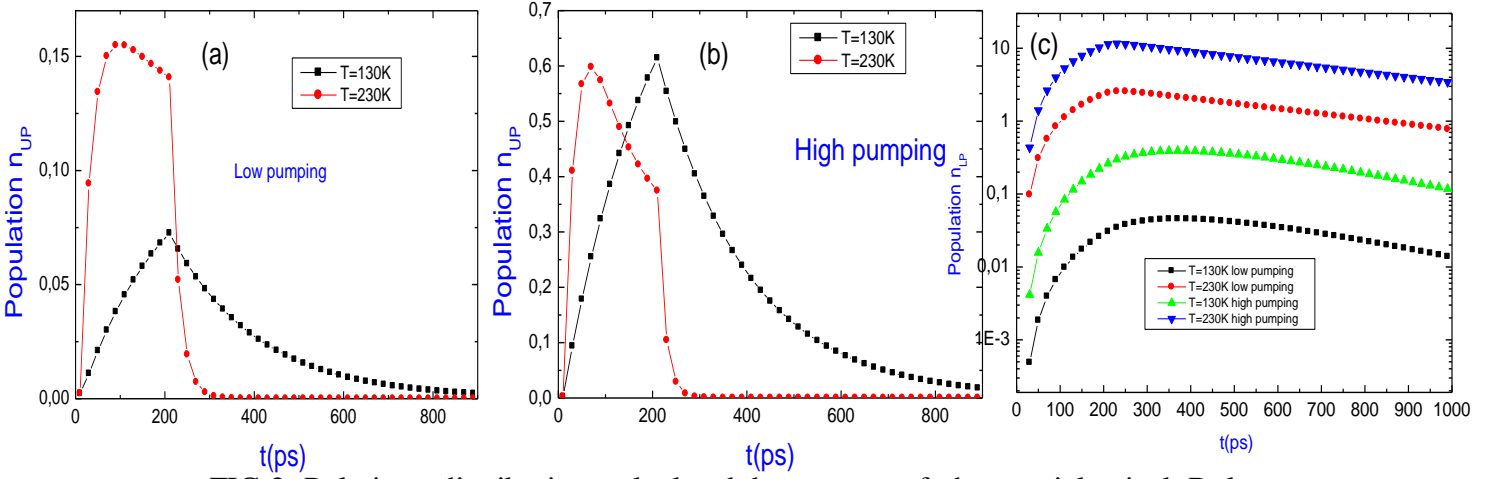

FIG.3: Polariton distribution calculated by means of the semiclassical Boltzmann equation in the (a). (b) UP branch and (c) LP branch below and above threshold at 130 and 230 K for a detuning-16 and 11meV, respectively.

FIG.3 (a) shows the calculated the polariton distribution $n_{UP}(t)$ below threshold for two different temperatures 130 and 230K. Below threshold, the rise time is short at 230K with respect to 130K. The population of the UP branch at 130K shows the presence of peak at the end of the 200ps of the pulse duration, which is moved towards 90ps as the temperature is increased. The shift as a function of the temperature (FIG. 3(a)) is attributed mainly to the decrease of the scattering time into the LP branch between the states **k** and **k′** (defined in Eq.2) as temperature is increased. In the high pumping regime, the polariton distribution $n_{UP}$ shows that the rise time become very shorter and the dynamics become much faster, as shown in FIG.3(b). At the end of the exciting pulse, FIG.3(b) shows that the decay time of $n_{UP}$ for temperature 130K is quite long in comparison with 230K. After 200ps, FIGs.3 (a) and (b) show that the UP occupation starts to decrease, which is related to the reduction of $n_{UP}(t)$ for increasing occupation numbers in the LP branch, as depicted in FIG.3(c). The evolution of the occupancy in the LP branch is represented in FIG.3(c), we note this distribution can reach a large occupation numbers when the pump rate is increased. Furthermore, the occupation of the LP mode increases rapidly versus temperature. We deduce that this enhancement is more effective at higher temperatures due to the scattering process by LO phonons. These characteristics are well reproduced by simulations based on the solution of semi-classical Boltzmann equations integrated over the time, which are presented in FIG.4. These simulations show the polariton distribution $n_{UP,\mathbf{k}=\mathbf{0}}$ and $n_{LP,\mathbf{k}=\mathbf{0}}$ versus pumping rate for Δ= -16 meV, T=130K and Δ=11 meV, T=230K, respectively.

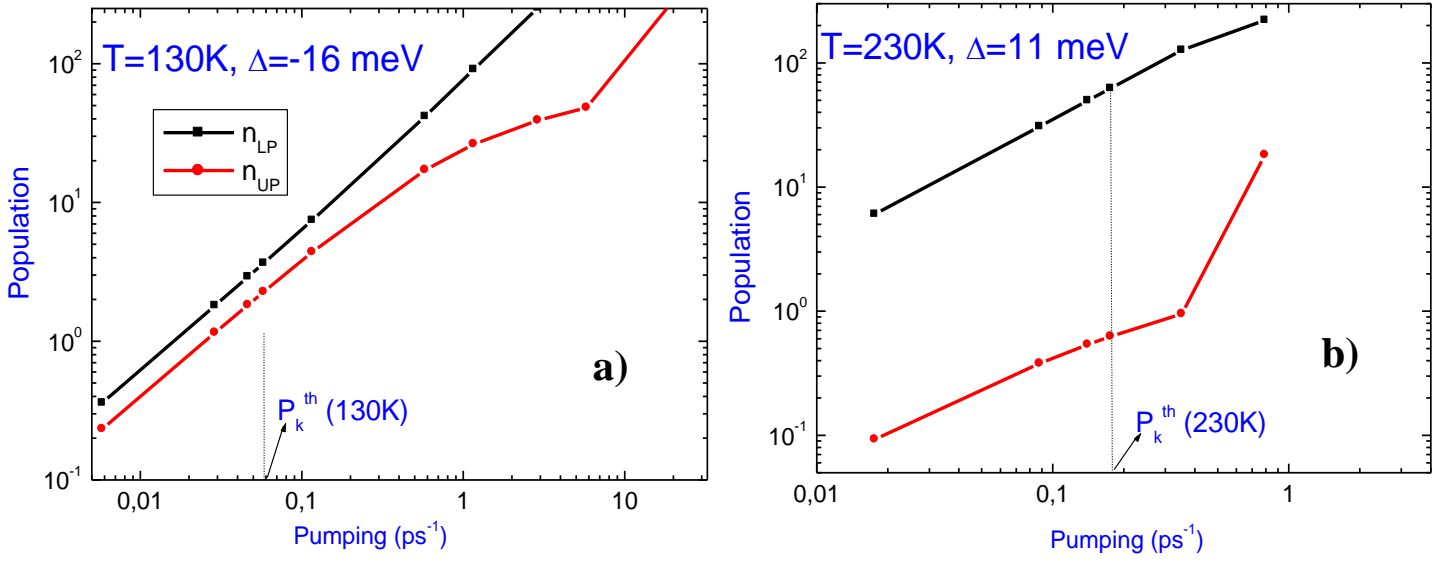

FIG.4: The population $n_{UP,k=0}$ and $n_{LP,k=0}$ as a function of optical pumping rate calculated using semi-classical Boltzmann equations for a detuning of -16 meV at T=130K and 11meV at T=230K.

As seen in these FIG.4 (a) and (b) below the threshold, both curves $n_{UP,k=0}$ and $n_{LP,k=0}$ are linear and directly proportional to the pump rate $P_k$; $n_{LP,k=0}$ becomes close to 1 while $n_{UP,k=0}$ remains small. When going toward a photonlike polariton, interaction with phonons becomes less efficient. Thus, for small pumping rate the LP branch at Δ= -16 meV (T=130K) is less occupied with respect to 11meV (T=230K). When the pumping increases, the UP population rises quadratically as reported in FIG.4(b), and an efficient polariton relaxation toward the LP state is obtained. The nonlinear behavior at higher pumping rate accelerate LP relaxation, as demonstrated by the time evolution of the lower state LPs in FIG.4(b). As seen in FIG.2 (a) and (b), the relaxation mechanism here is however different from the negative detuning case.

### V. Conclusion:

In conclusion, we have shown the formation of microcavity polaritons in a two dimensional atomic crystal of $WS_2$ embedded inside a microcavity over a temperature range 110–230 K, when the emergence of dark exciton and trion is neglected. A Rabi splitting of 39,98 meV is observed from the $WS_2$ microcavity owing to the coupling between the 2D excitons and the cavity photons. With the large binding energy and strong temperature dependence, the exciton energy shifts from 2.069 (130K) to 2.042eV (230K), creating cavity detuning varies from negative to positive values; which causes the corresponding changes in polariton dispersions. The tuning of polarition dispersion, lifetimes over a large temperature range enables the control of polariton emission dynamics. Our results can be understood in terms of Boltzmann equations in a two-state model under nonresonant excitation taking into account polariton scattering with optical phonons. These numerical simulations give an overall description of the polaritons dynamics, which depends on several parameters, the excitation of the system, radiative recombination rate and scattering processes. The results imply that at negative detunings, the relaxation kinetics is slow due to less efficient scattering processes, thus increasing the threshold pump rate needed to reach a polariton occupancy close to unity. On the other hand, for positive detunings, the threshold increases with temperature, because polariton relaxation becomes more efficient. As a result, occupation factors larger than unity are reached in the lowest **k** states.

# SUPPLEMENTARY INFORMATION

## Appendix A
### Exciton Binding Energy in Monolayer $WS_2$.

In this section, we theoretically investigate the properties of excitons in monolayer $WS_2$ deposited on HSQ/DBR substrate. Within the effective mass approximation, the neutral excitonis described by the following Hamiltonian $H = H_{CM} + H_r$. The First term $H_{CM} = \frac{\hbar^2 K^2}{2M_{ex}}$ describe the free motion of the center of mass. Their corresponding eigen function is $\frac{1}{\sqrt{S}} e^{iKR}$, where the in-plane center of mass wavevector $\mathbf{k}$ is a constant of motion and $S$ is the two-dimensional quantization area in the monolayer plane. $\mathbf{R} = \frac{m_e r_e + m_h r_h}{M_X}$ and $M_X = m_e + m_h$ are the exciton center of mass coordinates and mass, respectively. The following parameters have been considered for the calculations, effective electron and heavy hole masses $m_e = m_h = 0.34\, m_0$, with $m_0$ the free electron mass. The relative motion of the exciton can be described by the Hamiltonian $H_r = -\frac{\hbar^2 \nabla_r^2}{2\mu} + V_{2D}(r)$, where $\mu = \frac{m_e m_h}{m_e + m_h}$ is the reduced effective mass in the 2D plan and $\nabla_r$ is the gradient operator acting on the relative coordinate parallel to the monolayer plane is given by $r = \mathbf{r_e} - \mathbf{r_h}$. In the case of an excitonic system lying in a 2D plane, the interaction potential $V_{2D}(\mathbf{r})$ is given by the 2D screened electrostatic interaction potential derived by Keldysh [36, 37].

$$V_{2D}(\mathbf{r}) = -\frac{\pi e^2}{2\varepsilon_{eff}\, r_0}\left[H_0\left(\frac{r}{r_0}\right) - Y_0\left(\frac{r}{r_0}\right)\right]$$

$\varepsilon_{eff}$ is an environment-dependent dielectric constant. In the experiment, the monolayer $WS_2$ material is surrounded by an environment with dielectric constants $\varepsilon_1$ (top) and $\varepsilon_2$ (down).
$H_0$ and $Y_0$ are Struve and Bessel functions. The screening length $r_0$ can be related to the 2D polarizability of the monolayer material. In order to determine the eigenvalue equation, we use a wave function expansion technique [36]. We can calculate the eigenfunctions and eigenvalues of the system obtained numerically by a direct diagonalization of the full matrix resulting from the projection of the Hamiltonian. We find an exciton binding energy $E_b = 205\, meV$. Thus, the exciton energies are given by $E_{ex} = E_g(T) + \frac{\hbar^2 \mathbf{k}^2}{2M_X} - E_b$, where $E_g = 2.29\, eV$ is the band gap.

# Appendix B
# Modeling the temperature dependence of the $WS_2$ exciton energy

First, we investigate the temperature dependence of the exciton energy. The exciton energies redshift with increasing temperature, as shown in FIG. 1(b) and (d). It is described by the standard temperature dependence of semiconductor bandgaps as follows [46]:

$$E_g(T) = E_g - S_c <\hbar\omega> \left[\coth(\frac{<\hbar\omega>}{2k_BT}) - 1\right]$$

Where $S_c$ is a dimensionless coupling constant linked to electron-phonon coupling, $\hbar\omega$ is the average phonon energy, $k_B$ is the Boltzmann constant and T is the temperature.

In order to demonstrate the coherently coupled exciton polariton, the emergence of the trions and dark excitons must be neglected. So, we present here a more quantitative analysis of the trion population for comparison to that of the neutral bright exciton. First, we write the reaction rate equation for both bright and dark charged excitons formation $X_B + e \leftrightarrow X_B^t$ and $X_D + e \leftrightarrow X_D^t$, which are formed in the presence of an initial electron doping density. On the basis of the mass action law, the equation relating the concentrations of trions and excitons and the charge carrier density in TMDs can be written as follows [36,47]:

$$\frac{n_X n_e}{n_t} = 4\frac{Y_t^2}{Y_e^2 Y_X^2} exp\frac{-E_t}{K_BT}$$

$$\frac{n_h n_e}{n_X} = \frac{Y_X^2}{Y_e^2 Y_h^2} exp\frac{-E_b}{K_BT}$$

$Y_c = \frac{\hbar}{(\frac{2\pi}{m_c K_B T})^{1/2}}$ $(c = e, h)$ are the thermal wavelengths, and the prefactors correspond to a completely spin-equilibrated system. where $\hbar$ is the reduced Planck's constant, $K_B$ is the Boltzmann constant, T is the temperature, $E_t$ the trion binding energy is about 30meV [48] and $E_b$ is the exciton binding energy. From charge conservation and for the steady state conditions, we obtain:

$$n_e + n_X + 2n_t = n_p + n_B$$

$$n_p = n_X + n_t$$

$n_t$ and $n_e$ are the concentration of trion and free electrons. The exciton density $n_X$ is the sum of bright and dark exciton, the number of photoexcited electrons is given by $n_p$, and $n_B$ denotes the background electrons (doping level) before light excitation.

In order to determine the required trion and exciton densities, we have to solve the above equations. Thus, we can write

$$n_t = \frac{n_p + n_B + n_A - \sqrt{(n_p + n_B + n_A)^2 - 4n_p n_B}}{2}$$

The temperature dependent equilibrium constant is given by $n_A(T) = \frac{4M_{ex} m_e}{\pi \hbar^2 M_t} k_B T exp \frac{-E_t}{K_B T}$, $M_{ex}$ and $M_t = 2m_e + m_h$ are the exciton and trion effective masses respectively.

We plot in Figs.1 the bright trion and exciton densities as a function of temperature lies between 25 and 300K. Using the Saha equation to determine the charge density required and to explain the observed decrease in the neutral exciton population; as a consequence the creation of trions at low temperature. With decreasing temperature, Xiao-Xiao Zhang et al. [3] show the depletion of the neutral exciton population is due to the defect states. The emission intensity from the defect states in the PL spectra is large at low temperature (< 60 K). This behavior confirms the much longer emission time for the defect states compared to the neutral exciton. Hichri A. et al demonstrate the decrease of the bright exciton population when temperature decrease is due to the presence of dark exciton state [36]. From the obtained results, we note the emergence of trions and dark excitons at low temperature. Also, we clearly note the exchange of population between exciton and trion. This behavior confirms that the exciton (trion) formation is favored at high (low) temperature [3, 36]. Therefore, our studies are limited at $T > 130K$, where the exciton peak dominates.

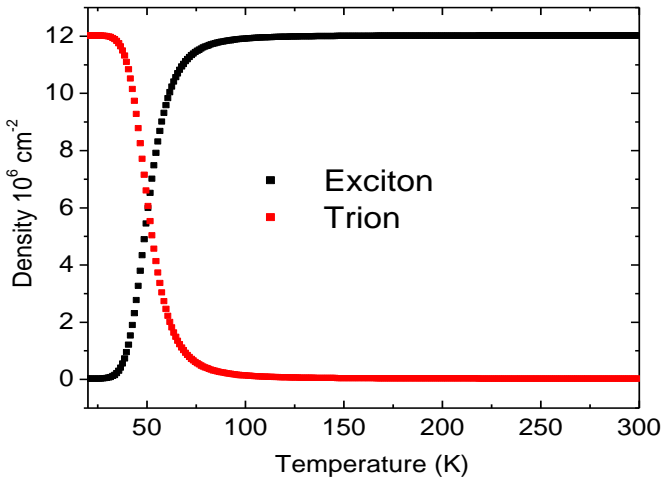

FIG.1. Temperature dependence of the population of neutral and charged excitons in monolayer $WS_2$, for fixed areal density of photoexcited electrons $n_p = 12\,10^6 cm^{-2}$ and pump to background ratio $s = \frac{n_p}{n_B} = 0.05$

# Appendix C
# The relaxation rate of exciton mediated by acoustic phonon
# &
# Rabi splitting between the UP and LP branches

The scattering rate of exciton by acoustic phonon (LA) from energy $E$ to all other states is calculated using the Fermi Golden rule [27].

$$W^{LA}_{ex,\mathbf{k}\to ex,\mathbf{k'}} = \frac{2\Pi}{\hbar}\sum_{\mathbf{q}}|\langle\Psi_{ex,\mathbf{k'}},0_{\mathbf{q}}|H_{ex\to LA}|\Psi_{ex,\mathbf{k}},1_{\mathbf{q}}\rangle|^2 \delta(E_{exc}(\mathbf{k'}) - E_{exc}(\mathbf{k}) + E_{LA}(\mathbf{q}))(N+1)$$

Here $0_{\mathbf{q}}$ and $1_{\mathbf{q}}$ are the initial and final phonons states. The phonons are taken to be the bulk phonons with a wave vector $\mathbf{q} = \mathbf{q}_{//} + \mathbf{q}_z$. Only the in-plane vector component is conserved. Accordingly, the exciton is represented by a quasi-two dimensional wavefunction $\psi_{ex} = \frac{1}{\sqrt{S}}e^{i\mathbf{kR}}\varphi_{1s}(\mathbf{r})\psi_e(z_e)\psi_h(z_h)$ [49]. The electron and hole wave functions $(\psi_e(z_e), \psi_h(z_h))$ are given by $\sqrt{\frac{2}{L}}\cos\frac{\pi z_i}{L}$ (i = e, h). The term $\sqrt{\frac{2}{L}}$ denotes the normalization constant and L is the average displacement of electrons and holes in the z direction perpendicular to the monolayer. $|\Psi_{ex,\mathbf{k}}\rangle$ is the exciton state with the in plane wave vector $\mathbf{k}$ and $E_{exc}(\mathbf{k})$ is the exciton energy. At a given temperature, the average phonon occupation number is given by the Bose distribution $N = \frac{1}{[\exp(\frac{E_{LA}(\mathbf{q})}{k_B T})-1]}$ with energy $E_{LA}(\mathbf{q}) = \hbar C_s \mathbf{q}$.

The relevant interaction between excitons and longitudinal-acoustic phonons is provided by the deformation-potential coupling with the electron and hole $D_e$ and $D_h$, respectively

$$H_{ex-LA} = \sum_{\mathbf{q}}[(\alpha_e(\mathbf{q})e^{i\mathbf{q}r_e} + \alpha_h(\mathbf{q})e^{i\mathbf{q}r_h})a_{\mathbf{q}}^+ + \text{c.c}]$$

The phonon creation operator is denoted by $a_{\mathbf{q}}^+$ in the mode $\mathbf{q}$, the term $|\alpha_i(\mathbf{q})|^2 = \frac{D_i^2}{2\rho V C_s^2}(\hbar C_s q)$, (i = e, h), where $\rho$ is the density, $C_s$ is the sound velocity and V is the volume of the crystal. The following parameters have been considered for the calculations [50] $C_s = 3.3 10^5 ms^{-1}, \rho = 2.4\ 10^{-6} kg/m^2, D_e = 5.08 eV, D_h = 3.98 eV$.

Evaluating the exciton-phonon interaction matrix element with the wavefunction of exciton, one finds [27],

$$|\langle\Psi_{ex,\mathbf{k'}},0_{\mathbf{q}}|H_{ex\to LA}|\Psi_{ex,\mathbf{k}},1_{\mathbf{q}}\rangle| = 2|\frac{(\alpha_e(\mathbf{q})I_e(q_z) + \alpha_h(\mathbf{q})I_h(q_z))}{I_e(\mathbf{q}_{//})}|\delta(\mathbf{q}_{//},\mathbf{k'}-\mathbf{k})$$

The superposition integrals for the electron (e) or hole (h) are given by

$$I_{e(h)}(q_z) = \frac{8\pi^2 \sin\frac{q_z L}{2}}{q_z L(4\pi^2 - q_z^2 L^2)} \quad and \quad I_{e(h)}(q_{//}) = \left[1 + \left(\frac{a_b m_{e(h)} q_{//}}{8(m_e + m_h)}\right)^2\right]^{-\frac{3}{2}}$$

The delta function conserves the momentum of the scattered exciton and phono, where $|\mathbf{q}_{//}| = |\mathbf{k}' - \mathbf{k}| = [(\mathbf{k}^2 + \mathbf{k}'^2 - 2\mathbf{k}\mathbf{k}'\cos\theta)]^{\frac{1}{2}}$, the angle between the exciton wave vector $\mathbf{k}'$ and $\mathbf{k}'$ is denoted by $\theta$. These integrals cut off the sums for $a_b^{-1} < q_{//}$ (where $a_b$ is the 2D Bohr radius) and for $q_z > \frac{2\pi}{L}$.

With this interaction matrix element and the sum over $q_z$, we obtain:

$$W_{exc,\mathbf{k}\to exc,\mathbf{k}'}^{LA} = \frac{1}{\hbar^3 \rho C_s^4} \int \frac{E_{exc}(\mathbf{k}') - E_{exc}(\mathbf{k})}{|q_{z_0}|} \left(\frac{[D_e I_e(q_{z_0}) + D_h I_h(q_{z_0})]^2}{\left[1 + \left(\frac{a_b m_e |\mathbf{k}'-\mathbf{k}|}{8(m_e+m_h)}\right)^2\right]^3}\right) [N+1] \, d\mathbf{k}' d\theta$$

The root of the argument of the delta function is given by $q_{z_0} = \sqrt{\frac{E_{exc}(\mathbf{k}') - E_{exc}(\mathbf{k})}{\hbar^2 C_s^2} - |\mathbf{k}' - \mathbf{k}|^2}$, we show that solutions of this equation do not exist for arbitrary value of $\mathbf{k}$ and $\mathbf{k}'$. The integration is taken over the range in which terms under the square root are non-negative. The above equation can be further simplified by assuming the exact two dimensional case, because the adimensional integrals $I_e(q_{z_0})$ and $I_h(q_{z_0})$ becomes equal to 1 when $q_{z_0} L$ tends to 0.

Thus the phonon-scattering rate $\Gamma_{ex}$ is given by:

$$\Gamma_{ex} = \sum_{\mathbf{k}'} W_{ex,\mathbf{k}\to ex,\mathbf{k}'}^{LA}$$

From the last equation we can deduce $\hbar\Gamma_{ex} = 15.7$ meV (130K) and 17.2 meV (230K). The measured reflectivity of microcavity without 2D semiconductor obtained by Xiaoze et al. [5], shows that the cavity resonance is achieved at 2.052eV with a linewidth $\hbar\Gamma_{cav} = 5$meV. Thus, the interaction potential $V_A$ in our model is given by [51]

$$V_A = \frac{\sqrt{\left(16\sqrt{\frac{E_g(T)E_b}{E_g - E_b}}\frac{e^2}{L_{eff}}\frac{(1-m_e)m_h}{2n_{eff}^2(m_e+m_h)}\right)^2 + (\hbar\Gamma_{ex} - \hbar\Gamma_{cav})^2}}{2}$$

This leads to aninteraction potential $V_A$ varies between 19 and 18 meV at temperature 130 and 230K. The interaction potentials clearly indicate presence of strongly coupled polariton states at all temperatures, with a Rabi splitting of $\hbar\Omega_{rabi} = 2\sqrt{(V_A)^2 - \frac{1}{4}(\hbar\Gamma_{ex} - \hbar\Gamma_{cav})^2} = 39.98$ meV at 130K.

**Appendix D**

**Emission of longitudinal-optical phonons (LO) within the microcavity polariton**

The scattering rate by LO phonons is given by the Fröhlich interaction between electron-phonon and hole-phonon. This transition rate is calculated in a similar way and contain a term $(N + 1)$ given by $= \frac{1}{[\exp(\frac{E_{LO(q)}}{k_B T})-1]}$ with energy $E_{LO}(q) = \hbar\omega_{LO} - bq^2$, $b = 14 \, 10^{-8} eV \mu m^2$ is a fitting parameter obtained from the LO phonon dispersion curve. In our case of monolayer $WS_2$, the LO phonon energy is calculated to be 43meV [52]. Here the relevant interaction between excitons and LO is provided by the Frôlich interaction. The Frôlich interaction constant of LO phonon with charge carriers in two-dimensional materials is given by $|\alpha(q)|^2 = \frac{e^2 \hbar \omega_{LO}}{2\varepsilon_0 V q^2}\left(\frac{1}{\varepsilon_\infty} - \frac{1}{\varepsilon_s}\right)$. The low-frequency and high-frequency relative dielectric constants are given by $\varepsilon_s$ and $\varepsilon_\infty$, respectively, V is the volume of the crystal.

Evaluating the interaction matrix element we can write:

$$|\langle UP_k, 0_q|H_{ex \to LO}|LP_{k'}, 1_q\rangle| = 2\alpha(q) X_k X_{k'} \left|\frac{(I_e(q_z) + I_h(q_z))}{I_e(q_{//})}\right| \delta(\mathbf{q}_{//}, \mathbf{k'} - \mathbf{k})$$

where $X_\mathbf{k}$ ($X_{\mathbf{k'}}$) are the Hopfield coefficients whose square modules gives the exciton content in the polariton state at wave vector $\mathbf{k}(\mathbf{k'})$, respectively. The eigenstates of the coupled system exciton-photon are given by $|UP_\mathbf{k}>$ and $|LP_{\mathbf{k'}}>$.

The integration with the delta function is the same as for acoustic phonons. The explicit expression of $W^{LO}_{UP,\mathbf{k} \to LP,\mathbf{k'}}$ can be written as follows:

$$W^{LO}_{UP,\mathbf{k} \to LP,\mathbf{k'}} = \frac{e^2 \hbar \omega_{LO}}{2\hbar\varepsilon_0}\left(\frac{1}{\varepsilon_\infty} - \frac{1}{\varepsilon_s}\right)\frac{|X_\mathbf{k}|^2 |X_{\mathbf{k'}}|^2}{q_{z_0}\Delta E}\left(\frac{[I_e(q_{z_0}) + I_h(q_{z_0})]^2}{\left[1 + \left(\frac{a_b m_e |\mathbf{k'}-\mathbf{k}|}{8(m_e+m_h)}\right)^2\right]^3}\right)[N+1]$$

Where the root of the argument of the delta function is given by $q_{z_0} = \sqrt{\frac{\Delta E}{b} - |\mathbf{k'} - \mathbf{k}|^2}$ and $\Delta E = \hbar\omega_{LO} + (E_{LP}(\mathbf{k'}) - E_{UP}(\mathbf{k}))$. Then, we need to specify the energy dispersion of the polaritons modes. We use the results which have been presented in section II.

The phonon-scattering time $\tau_{LO}$ is given by $\frac{1}{\tau_{LO}} = \sum_{\mathbf{k'}} W^{LO}_{UP,\mathbf{k} \to LP,\mathbf{k'}}$.

**References:**

[1] Q. Wang, L. Sun, Bo Zhang, Ch. Chen, X. Shen and W. Lu, J. Opt. Express, 24, 7, 7152 (2016)

[2] L. Zhang, R. Gogna, W. Burg, E. Tutuc and H. Deng, J. Nat. Comm. 9,713 (2018)

[3] X. Liu, W. Bao, Q. Li, Ch. Ropp, Y. Wang, and X. Zhang, J. Phys. Rev. Lett. 119, 027403 (2017)

[4] R. Tao, K. Kamide, M. Arita, S. Kako, and Y. Arakawa, J ACS Photonics, 3 (7) 1182–1187, (2016)

[5] F. Scafirimuto, D. Urbonas, U. Scherf, R. F. Mahrt, and Thilo Stöferle ,J. ACS Photonics, 5 (1) 85–89, (2018)

[6] L. C. Flatten, Z. He, D. M. Coles, A.A. P.Trichet, A.W. Powell, R.A.Taylor, J. H.Warner and J. M. Smith, Scientific reports 6, 33134 (2016)

[7] M. Richard, J. Kasprzak, R. André, R. Romestain, L. S. Dang, G. Malpuech, and A. Kavokin, J. Phys. Rev. B 72, 201301 (2005)
J. Kasprzak, M. Richard, S. Kundermann, A. Baas, P. Jeambrun, J. M. J. Keeling, F. M. Marchetti, M. H. Szymańska, R. André, J. L. Staehli, V. Savona, P. B. Littlewood, B. Deveaud, and L. S. Dang, J. Nature 443, 409 (2006)

[8] R. Balili, V. Hartwell, D. Snoke, L. Pfeiffer, and K. West, J. Science 316, 1007 (2007)

[9] D. Bajoni, P. Senellart, E. Wertz, I. Sagnes, A. Miard, A. Lemaître, and J. Bloch, J. Phys. Rev. Lett. 100, 047401 (2008)

[10] G. Malpuech, A. Di Carlo, A. Kavokin, J. J. Baumberg, M. Zamfirescu, and P. Lugli, J.Appl. Phys. Lett. 81, 412 (2002).

[11] M. Zamfirescu, A. Kavokin, B. Gil, G. Malpuech, and M. Kaliteevski, J. Phys. Rev. B 65, 161205 (2002).

[12] T. Ch. Lu, Y.Yu Lai, Y. Pin Lan, S.Wei Huang, J.-Rong Chen, Y.Chi Wu, W. Feng Hsieh, and H. Deng. J. Opt. Express, 20(5):5530-5537 (2012).

[13] S Kena-Cohen, M. Davanco, and S. Forrest. J. Phys. Rev. Lett. 101(11), 116401, (2008).

[14] S Kena-Cohen and S. R. Forrest. J. Nat. Photonics, 4, 371-375 (2010).

[15] X. Liu, T. Galfsky, Z. Sun, F. Xia, E. Lin, Y.H. Lee, S. Kéna-Cohen, and V. M. Menon, J. Nat. Photonics 9(1), 30–34 (2014).

[16] S. Dufferwiel, S. Schwarz, F. Withers, A. A. P. Trichet, F. Li, M. Sich, O. Del Pozo-Z amudio, C. Clark, A. Nalitov, D. D. Solnyshkov, G. Malpuech, K. S. Novoselov, J. M.

Smith, M. S. Skolnick, D. N. Krizhanovskii, and A. I. Tartakovskii, J. Nat. Commun. 6, 8579 (2015).

[17] G. Wang, C. Robert, A. Suslu, B. Chen, S. Yang, S. Alamdari, I. C. Gerber, T. Amand, X. Marie, S. Tongay and B. Urbaszek, J. Nat. Comm. 6, 10110 (2015)

[18] K. G. Lagoudakis, F. Manni, B. Pietka, M. Wouters, T. C. H. Liew, V. Savona, A. V. Kavokin, R. Andr e, and B. Deveaud Pledran, J. Phys. Rev. Lett. 106, 115301 (2011).

[19] G. Christmann, G. Tosi, N. G. Berlo_, P. Tsotsis, P. S. Eldridge, Z. Hatzopoulos, P. G. Savvidis, and J. J. Baumberg, J. Phys. Rev. B 85, 2 35303 (2012).

[20] F. Manni, K. G. Lagoudakis, T. C. H. Liew, R. Andr_e, and B. Deveaud Pledran, J. Phys. Rev. Lett. 107, 106401 (2011).

[21] J. Kasprzak, D. D. Solnyshkov, R. Andre, Le Si Dang, and G. Malpuech, J. Phys. Rev. Lett. 101, 146404 (2008).

[22] D. Porras, C. Ciuti, J. J. Baumberg, and C. Tejedor, J. Phys. Rev. B 66, 085304 (2002).

[23] A. Amo, D. Sanvitto, F. P. Laussy, D. Ballarini, E. del Valle, M. D. Martin, A. Lemaitre, J. Bloch, D. N. Krizhanovskii, M. S. Skolnick, C. Tejedor, and L. Vina, J. Nature 457, 291 (2009).

[24] T. D. Doan, H. T. Cao, D. B. Tran Thoai, and H. Haug, J. Phys. Rev. B 72, 085301 (2005).

[25] H. T. Cao, T. D. Doan, D. B. Tran Thoai, and H. Haug, J. Phys. Rev. B 77, 075320 (2008).

[26] F. Tassone and Y. Yamamoto, J. Phys. Rev. B 59, 10830 (1999).

[27] F. Tassone, C. Piermarocchi, V. Savona, A. Quattropani, and P. Schwendimann, J. Phys. Rev. B 56, 7554 (1997)

[28] G. Malpuech, A. Kavokin, A. Di Carlo, and J. J. Baumberg J. Phys . Rev. B 65, 153310 (2002)

[29] R. P. Stanley, S. Pau, U. Oesterle, R. Houdre, and M. Ilegems, J. Phys. Rev . B 55, 8 (1997)

[30] M. Maragkou, A. J. D. Grundy, T. Ostatnický, and P. G. Lagoudakis, J. Appl. Phys. Lett. 97, 111110 (2010)

[31] A. Imamoglu, R. J. Ram, S. Pau, and Y. Yamamoto, J. Phys. Rev. A 53, 4250 (1996).

[32] S. Pau, G. Björk, H. Cao, F. Tassone, R. Huang, Y. Yamamoto, and R. P Stanley, Phys. Rev. B 55, R1942 (1997)

[33] F. Tassone, C. Piermarocchi, V. Savona , A. Quattropani , P. Schwendimann, J. Phys. Rev. B 53, R7642– 7645 (1996)

[34] L. Orosz, F. Reveret F. Medard, P. Disseix, J. Leymarie, M. Mihailovic, D. Solnyshkov, G. Malpuech, J. Zuniga-Perez, F. Semond, M. Leroux, S. Bouchoule, X. Lafosse, M. Mexis, C. Brimont, and T. Guillet, J. Phys. Rev. B. 85, 121201(R) (2012)

[35] O. Jamadi, F. Reveret, E. Mallet, P. Disseix, F. Medard, M. Mihailovic, D. Solnyshkov, G. Malpuech, J. Leymarie, X. Lafosse, S. Bouchoule, F. Li, M. Leroux, F. Semond, and J. Zuniga-Perez, J.Phys. Rev. B 93, 115205 (2016)

[36] A. Hichri, I. Ben Amara, S. Ayari, and S. Jaziri, J. App. Phys. 121, 235702 (2017)

[37] A. Chernikov, T. C. Berkelbach, Heather M. Hill, Albert Rigosi, Yilei Li, O. B. Aslan, David R. Reichman, Mark S. Hybertsen, and Tony F. Heinz, J. Phys . Rev . Lett PRL 113, 076802 (2014)

[38] M. Palummo, M. Bernardi and J. C. Grossman, J. Nano Lett., 15 (5), 2794–2800(2015)

[39] H. Wang, Ch. Zhang, W. Chan, Ch. Manolatou, S. Tiwari, and F. Rana, J. Phys. Rev . B 93, 045407 (2016)

[40] L. C. Andreani, F. Tassone, and F. Bassani, J. Sol. State Commun. 77, 641 (1991).

[41] N. B. Mohamed, H. En Lim, F. Wang, S. Koirala, S. Mouri, K. Shinokita, Y. Miyauchi, and K. Matsuda, J. Appl. Phys. Express 11, 015201 (2018)

[42] M. Muller and Joel Bleuse, J. Phys. Rev. B 62, 24 (2000)

[43] P. Senellart and J. Bloch, J. Phys . Rev. Lett. 82, 6 (1999)

[44] J. Levrat, R. Butté, E. Feltin, J. F. Carlin, and N. Grandjean, J. Phys. Rev B 81, 125305 (2010)

[45] G. V. Kolmakov, L. M. Pomirchi, and R. Ya. Kezerashvili, J. Optic. Society of America B 33, C72-C79 (2016)

[46] J. Cuadra, Denis G. Baranov, M. Wersäll , R. Verre , Tomasz J. Antosiewicz , and T. Shegai, J. Nano. Lett, 18 (3), 1777–1785( 2018)

[47] A. Esser, E. Runge, R. Zimmermann, and W. Langbein, J. Phys. Stat. Sol. (a) 178, 489 (2000)

[48]   M. Szyniszewski, E. Mostaani, N. D. Drummond, V. I. Fal'ko, J. Phys. Rev. B 95, 081301(R) (2017)

[49]   A. Thilagam, J. Appl. Phys. 120, 124306 (2016)

[50]   A. Thilagam, J. Appl. Phys. 119, 164306 (2016)

[51]   N Lundt, A Maryński, E Cherotchenko, A Pant, X Fan, S Tongay, G Sęk, A V Kavokin, S Höfling and C Schneider , J. 2D Mater. 4 015006 (2017)

[52]   Zh. Jin, X. Li, Jeffrey T. Mullen, and Ki Wook Kim, J. Phys. Rev 90, 045422 (2014)